\documentclass[12pt]{iopart}
\usepackage{iopams}
\usepackage{setstack}
\usepackage[latin1]{inputenc}
\usepackage[T1]{fontenc}
\usepackage{graphicx}
\usepackage{graphics}
\usepackage{pstricks}
\usepackage{epsfig}
\usepackage{subfigure}
\newcommand{\Frac}[2]{\frac{{\displaystyle #1}}{{\displaystyle #2}}}
\begin{document}
\def\theequation{\arabic{section}.\arabic{equation}}
\title{The phase space view of $f(R)$ gravity}
\author{José C. C. de Souza$^{1,2}$ and Valerio 
Faraoni$^2$}
\address{$^1$ Departamento de Física Matemática, Instituto de
  Física, Universidade de São
  Paulo, C.P. 66318, 05315-970, São Paulo, SP, Brazil. }
\address{$^2$ Physics Department, Bishop's University, Sherbrooke,
Qu\`{e}bec, Canada J1M~0C8}
\ead{jsouza@ubishops.ca, vfaraoni@ubishops.ca}

\begin{abstract}
 We study the geometry of the phase space of spatially flat
 Friedmann-Lemaitre-Robertson-Walker models in $f(R)$ gravity, for a
 general form of the function $f(R)$. The equilibrium points (de
 Sitter spaces) and their stability are discussed, and a comparison
 is made with the phase space of the equivalent scalar-tensor
 theory. New effective Lagrangians and Hamiltonians are also 
presented.
\end{abstract} \vspace*{1truecm}
\maketitle

\def\theequation{\arabic{section}.\arabic{equation}}


\section{Introduction}
\setcounter{equation}{0}
\setcounter{page}{2}

The observation of type Ia supernovae \cite{SN} tells us that 
the
universe is accelerating, while cosmic microwave background
experiments \cite{CMB} point to a spatially flat universe. 
Within the context of
general relativity, the acceleration is explained by a form of 
dark
energy which accounts for 70\% of the cosmic energy content with
exotic equation of state $P\approx -\rho$. It is even possible 
that the effective equation of state of this dark
energy component has $P<-\rho$ (phantom energy) \cite{phantom}. 
As an
alternative to postulating this strange form of unseen energy, 
various
modifications of gravity have been proposed. Among them probably 
the
most popular is the so-called $f(R)$ gravity \cite{MG}, 
described by the action
\begin{equation}\label{action_1}
S=\int d^4 x \, \sqrt{-g} \,\,\frac{f(R)}{2} +S^{(matter)} \; ,
\end{equation}
\noindent where $R$ is the Ricci curvature, $f(R)$ is a 
non-linear function of $R$, $g$ is the determinant of
the metric tensor $g_{ab}$, and $8\pi G=1$ ($G$ being Newton's
constant) in our notations. Other notations follow
  Ref.~\cite{Wald}. $f(R)$ gravity comes in three versions: the
so-called {\em metric formalism}, in which the  action 
(\ref{action_1}) is
varied with respect to $g^{ab}$ and the field equations are of fourth
order; the {\em Palatini formalism} \cite{Palatini}, in which the
action (\ref{action_1}) is varied with respect to both the metric and
the connection, yielding second order equations, and in which the
matter part of the action $S^{(matter)}$ is independent of the
connection; and the so-called {\em metric-affine gravity} 
version in  which also $S^{(matter)}$ depends explicitly on it 
\cite{metricaffine}.  The physical motivation for all these 
versions of modified gravity lies in the desire to explain the 
cosmic acceleration without dark energy (and, especially, 
to get away from the embarassing notion of phantom energy).

Here we restrict the analysis to the metric formalism, in which 
the field equations are
\begin{equation}\label{fieldeq}
f'(R) R_{ab}-\frac{f(R)}{2}\, g_{ab}=\nabla_a\nabla_b f' -g_{ab} 
\Box f'+T_{ab} 
\end{equation}
(a prime denotes differentiation with respect to $R$).

While several models of metric $f(R)$ cosmology are found in the
literature, most works consider specific forms of $f(R)$ which are not
motivated by fundamental reasons, and are usually chosen on the basis
of simplicity or ease of calculation. It is more interesting not to
make assumptions on the form of $f(R)$ and try instead to understand
this class of theories in as general a way as possible, without
choosing the form of $f(R)$. In a previous paper we have shown that
theories with $f''<0$ are not viable due to instabilities in the Ricci
scalar \cite{mattmodgrav,SawickiHu}. Problems with the weak field
limit have also been pointed out for many choices of $f(R)$
\cite{weakfieldlimit}. Other issues to be studied include the correct
sequence of cosmological eras (inflation, radiation era, matter era,
present accelerated era) and whether smooth transitions between them
are possible \cite{Amendolaetal}; the presence of ghosts and
instabilities \cite{instabilities}; and the well-posedness of the
Cauchy problem \cite{mattmodgrav}.

Here we provide a description of the phase space for general metric
$f(R)$ cosmology, i.e., without specifying the form of the function
$f(R)$. We focus on the geometry of the phase space, the existence of
equilibrium points and their stability, and the phase space picture of
the equivalent scalar-tensor theory.

Further motivation for our study comes from scenarios of inflation in
the early universe employing quadratic corrections to the
Einstein-Hilbert Lagrangian, i.e., $f(R)=R+aR^2$
\cite{Starobinsky}. (The phase space of such theories has already
received attention in the literature
\cite{CapozzielloOcchioneroAmendola}.) Quadratic corrections are
motivated by attempts to renormalize Einstein's theory
\cite{renormalize}, and are included as special cases in the following
discussion.

It is well known that $f(R)$ gravity can be mapped into the 
Einstein conformal frame in which the theory is equivalent to a 
scalar field minimally coupled to the Ricci curvature 
(\cite{Whitt84}; see also \cite{MagnanoSokolowski,FGN}). 
Although this version is completely equivalent to ``Jordan 
frame'' $f(R)$ gravity, at least at the classical level, there 
have been much confusion and misunderstanding about the mapping 
into the Einstein frame (see, e.g., the recent discussion of 
\cite{FaraoniNadeau07}) and we prefer to proceed directly 
without using this conformal mapping.

The plan of this paper is as follows.  In
Sec.~\ref{f_R}, the geometry of the phase space of spatially 
flat Friedmann-Lemaitre-Robertson-Walker (FLRW) cosmology is
studied. In Sec.~\ref{equiv_stt}, the phase space view is 
compared with that of the equivalent scalar-tensor theory, and 
effective Lagrangians and Hamiltonians are presented for both 
cases. Sec.~\ref{conclusions} contains a discussion and the 
conclusions.

\section{The phase space of $f(R)$ gravity}\label{f_R}
\setcounter{equation}{0}

In this section, we study the geometry of the phase space of 
$f(R)$
gravity. Since we study regimes dominated by corrections to
Einstein gravity (for example, the late-time acceleration of the
universe, or the early epoch of inflation in Starobinsky-like scenarios
\cite{Starobinsky}), we omit the matter part of the action
(\ref{action_1}).  Motivated by the recent cosmological 
observations, we restrict the analysis to the spatially flat
FLRW line element
\begin{equation}\label{metric}
ds^2= -dt^2+ a^2(t)(dx^2+dy^2+dz^2)
\end{equation}
in comoving coordinates $\left( t, x, y, z \right)$. The field 
equations then reduce to
\begin{equation}\label{H_squa}
H^2=\frac{1}{3f'}\left[\frac{Rf'-f}{2}-3H\dot{R}f''\right] \;,
\end{equation}
\begin{equation}\label{H_dot}
2\dot{H}+3H^2=-\frac{1}{f'}\left[f'''(\dot{R})^2 
+2H\dot{R}f''+\ddot{R}f''+\frac{1}{2}(f-Rf')\right] \;,
\end{equation}
where an overdot denotes differentiation with
respect to $t$. We assume that $f'>0$ to 
have a positive effective gravitational
coupling; furthermore, modified gravity  theories suffer from 
violent instabilities unless
$f''\geqslant0$ (\cite{DolgovKawasaki, 
SO1, mattmodgrav,SawickiHu} --- 
see 
also \cite{BarrowOttewill}) and it seems appropriate to 
exclude 
the case $ f''<0$ on a physical basis. The field equations are 
of fourth order in $a(t)$. However, when the curvature index 
$k=0$, $a$ only appears in the combination $H\equiv \dot{a}/a$. 
Since the Hubble parameter $H$ is a  cosmological
observable, it is convenient to adopt it as the dynamical variable;
then, the field equations (\ref{H_squa}) and (\ref{H_dot}) are of
third order in $H$. The elimination of $a$ is not possible when 
the curvature index $k\neq  0$, or when a fluid with density 
$\rho=\rho(a) $ is included in the picture.  We can describe the 
dynamics in the 
three-dimensional
phase space $\left( H, R, \dot{R} \right) $, then the 
Hamiltonian constraint (\ref{H_squa}) implies that the orbits of 
the solutions lie on an energy surface $\Sigma$ and there is an 
expression for $\dot{R}$ for
any given value of the other two variables $ \left( H, R 
\right)$. This is given
explicitly by eq.~(\ref{H_squa}) as
\begin{equation}\label{R_dot}
\dot{R}(H, R)=\frac{Rf'-f-6f'H^2}{6Hf''} \; ,
\end{equation}
\noindent where we assume that $f''>0$. It is therefore possible to
eliminate the variable $\dot{R}$ given a pair $(H, R)$. Note that
there is only one value of $\dot{R}$ for each given value of $(H,
R)$. This situation is different from a general scalar-tensor theory
in which one obtains {\em two} values $\dot{\phi}_{\pm}(H, \phi)$ for
any value of the pair $(H, \phi)$ ($\phi$ being the scalar field of
gravitational origin) because the Hamiltonian constraint is a
quadratic equation in $\dot{\phi}$ in which the term $\dot{\phi}^2$ is
multiplied by the Brans-Dicke ``parameter'' $\omega(\phi)$ (which in
general scalar-tensor gravities becomes a function of
$\phi$ instead of the constant parameter appearing in Brans-Dicke
theory \cite{BransDicke, VFAnnPhys}). The scalar-tensor 
equivalent of $f(R)$
gravity has $\omega=0$ and no quadratic terms in $\dot{\phi}$ 
--- see the next section. This situation reflects the fact that 
$\dot{R}\left( H, R \right)$ is single-valued.

By eliminating the variable $\dot{R}$, the orbits of the 
solutions of eqs.~(\ref{H_squa}) and (\ref{H_dot}) are confined 
to the two-dimensional, curved, surface $\Sigma$ described by
eq.~(\ref{R_dot}) in the three-dimensional space $ \left( H, R, 
\dot{R} \right)$. By contrast, in general scalar-tensor theories 
with $\omega \ne 0$, $\dot{\phi}=\dot{\phi_\pm}(H, \phi)$ is a 
double-valued function and the corresponding surface 
$\Sigma_{\pm}$ is two-sheeted \cite{VFAnnPhys}. As a 
consequence of the two-dimensional nature of $\Sigma$, there can 
be no chaos in the dynamics. Although the standard
Poincaré-Bendixson theorem \cite{PoincareBendixson} only applies 
to a flat, compact
and simply connected two-dimensional phase space, it is rather
straightforward to prove the absence of chaos in the 
two-dimensional,
two-sheeted phase space of scalar-tensor gravity
\cite{FaraoniJensenTheuerkauf}, and the proof is easily extended to
the simpler one-sheeted phase space of $f(R)$ gravity.

As a consequence of the fact that the phase space is constituted 
of a
single sheet, the projections of the orbits onto the $(H, R)$ plane
can not intersect one another, contrary to what happens in the general
scalar-tensor case \cite{VFAnnPhys}. In general, there is a {\em
  forbidden region} in the phase space, corresponding to 
$H^2<0$ and here called $\mathcal{F}=\mathcal{F}_1 \cup 
\mathcal{F}_2$, where
\begin{equation}\label{forb_reg}
\mathcal{F}_1\equiv \left\{ \left(H, R \right): Rf'(R) 
-f(R)-6H\dot{R} \left(H, R \right) f''(R)<0 \right\} 
\end{equation}
and 
\begin{equation}
\mathcal{F}_2 \equiv \left\{ \left(H, R \right): f(R), f'(R), \; 
\mbox{or} \;\; f''(R) \;\; \mbox{are not defined}  \right\} \;.
\end{equation}
In addition, we assume that $f'>0$ to ensure a positive 
gravitational coupling and that $f''>0$ to avoid instabilities, 
which restricts further the region dynamically allowed to the 
orbits of the solutions. The boundary of the connected subset  
$\mathcal{F}_1$ of the  forbidden region is
\begin{equation}\label{bound}
\mathcal{B}\equiv \{(H, R, \dot{R}): f'(R)H^2=0\}\; ,
\end{equation}
which defines a curve in the surface $\Sigma$, along which the 
effective energy density $\rho_{eff}$ given by 
eq.~(\ref{density}) below vanishes.

A difference with respect to general scalar-tensor 
cosmology stands out: in these theories, the quadratic 
equation for the time derivative $\dot{\phi} $ of the  
Brans-Dicke-like scalar $\phi(t) $ 
provides two distinct roots $\dot{\phi}_{\pm}$ which differ by a 
square root term entering with positive or negative sign, and 
there is a forbidden region corresponding to the argument of this 
square root being negative \cite{VFAnnPhys}. As a result, the 
forbidden region intersects the two-sheeted energy surface 
$\Sigma$, creating some ``holes'' in this surface. The orbits of 
the solutions can change sheet at the boundary of these 
holes. In $f(R)$ cosmology instead, there is no such square root 
term and the forbidden region does not intersect the energy 
surface $\Sigma$.

Having chosen $\left( H, R \right) $ as variables, the 
equilibrium points are given by $(\dot{H},
\dot{R})=(0, 0)$ and are necessarily the de Sitter spaces 
\begin{equation}\label{desitter}
\left( H, R, \dot{R} \right)=\left( H_0, 12H_{0}^{2}, 
0 \right)= \left(\pm\sqrt{\frac{f_0}{6f'_0}}\,\,,
\frac{2f_0}{f'_0}\,\,, 0\right)\; ,
\end{equation}
where $f_0\equiv f(R_0)$, $f'_0\equiv f'(R_0)$, {\em etc.} These
fixed points only exist when the condition $f_0/f'_0 \geq 0$ is
satisfied. There is only one condition for the existence of the de
Sitter space, contrary to general scalar tensor gravity
\cite{VFAnnPhys} because eq.~(\ref{H_squa}) is not independent 
of eq.~(\ref{H_dot}) and there is only one independent equation 
left. According to the observational data available \cite{SN}, 
the present 
state of the universe is characterized by an effective equation 
of state $P \simeq -\rho$, i.e., if the $f(R)$ model applies, we 
are near a de Sitter fixed point.

If $f_0 \geqslant 0$ and 
$f'_0>0$, two de Sitter equilibrium points
(one contracting and one expanding) always exist in the $\dot{R}=0$
plane with $H_{0}^{(\pm )}=\pm\sqrt{\frac{f_0}{6f'_0}}$;  
they degenerate into a single point, a Minkowski space, if 
$f(R_0)=0$, in which case they lie on the boundary $\mathcal{B}$ 
of the forbidden region $\mathcal{F}_1$. The latter approaches 
the energy surface 
$\dot{R}=\dot{R}\left( H, R \right)$ at these Minkowski 
space points, which are the only possible global 
Minkowski solutions. 

The  stability of the de Sitter equilibrium points of
modified gravity with respect to inhomogeneous (space and
time-dependent) perturbations was studied in
Refs.~\cite{VFdSstability} by using the covariant and 
gauge-invariant
formalism of Bardeen-Ellis-Bruni-Hwang \cite{Bardeen}, in Hwang's
version valid for generalized gravity \cite{Hwang}. The result 
is the covariant and gauge-invariant condition for linear 
stability of de Sitter space with respect to inhomogeneous 
perturbations
\begin{equation}\label{stab} 
\frac{(f'_0)^2-2f_{0}f''_{0}}{f'_{0}f''_0}\geqslant 0 \;.
\end{equation} 
The contracting de Sitter spaces with $H_0 <0$ are always
unstable \cite{VFdSstability}. To linear order, expanding de 
Sitter spaces 
are always stable with respect to tensor modes\footnote{It turns 
out that (\ref{stab}) coincides with the stability condition of 
de Sitter space with respect to {\em
homogeneous} perturbations 
\cite{VFstability2, VFPRDBriefReport}.}. The 
equilibrium points $ \left( H, R,
\dot{R} \right)= \left( H_0, 12H^2_0, 0 \right) $ always lie in 
the $ \left( H, R \right) $ plane of the
phase space.

\begin{figure}[!htb]\centering
\epsfig{file=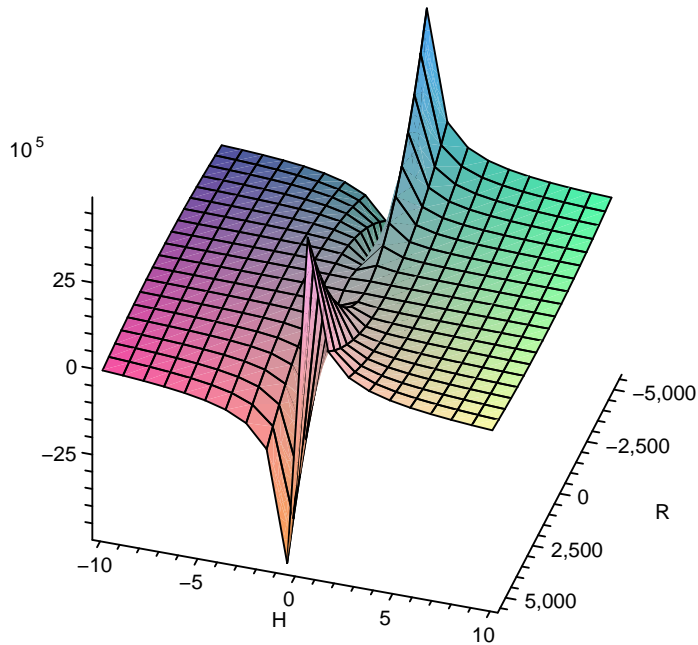,scale=0.7,angle=0}
 \caption{Phase space portrait for the model $f(R)=R+\alpha R^2$, with
$\alpha=0.0001$. The vertical axis shows $\dot{R}$ and $R$, $H$,
$\dot{R}$, $\alpha$ are measured in arbitrary units.}
\label{f_R_quad}
\end{figure}

\begin{figure}[!htb]\centering
\epsfig{file=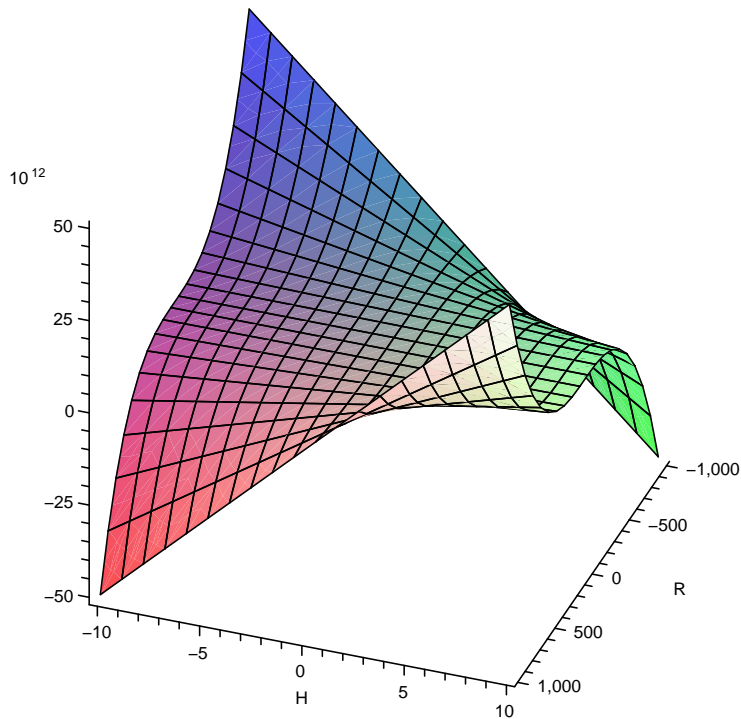,scale=0.8,angle=0}
 \caption{Phase space for the model $f(R)=R-\Frac{\mu^4}{R}$, with
$\mu^4=0.0001$, in arbitrary units, with $\dot{R}$ on the vertical
axis.}
\label{f_R_inv}
\end{figure}

Examples of the phase space geometry for two choices of the 
function $f(R)$ in the gravitational action (\ref{action_1}) are 
presented in fig.~\ref{f_R_quad} and fig.~\ref{f_R_inv}. 

It is useful to obtain the effective Lagrangian $L$ and 
Hamiltonian $E$ for $f(R)$ cosmology in FLRW space with $a$ and $R$ as
Lagrangian coordinates. To the best 
of our knowledge, the effective Lagrangian and Hamiltonian 
(\ref{lagrangian}) and (\ref{hamiltonian}) below have 
not been presented in the literature, and will be used in future 
work. For example, they can be used in the search for point-like 
symmetries (see Ref. \cite{pointlike}).
An analysis of the field equations leads one to conclude 
that the effective Lagrangian is
\begin{equation}\label{lagrangian}
L \left( a, R, \dot{a}, \dot{R} \right)=a^3\left[ 
6H^2f'+6Hf''\dot{R}+f'R-f 
\right]
\end{equation} 
and that the effective energy\footnote{The quantity $E$ can also be
  regarded as an effective Hamiltonian but it does not contain
  explicitly the generalized momenta.} is
\begin{equation}\label{hamiltonian}
E \left( a, R ,\dot{a},\dot{R} \right) =a^3 \left( 
6H^2f'+6Hf''\dot{R}-f'R+f 
\right) \;.
\end{equation}
The canonical momenta conjugated to the generalized coordinates $a$
and $R$ are
\begin{equation}
p_a=\frac{\partial L}{\partial \dot{a}}=6a^2 \left( 2Hf'+f'' 
\dot{R}\right) \;, \;\;\;\;\;\;\;\;\;
p_R=\frac{\partial L}{\partial \dot{R}}=6a^3  H f'' \;.
\end{equation}
The Hamiltonian constraint (\ref{H_squa}) corresponds to $E=0$. 
By choosing $a$ and $R$ as canonical variables, the 
Euler-Lagrange equation $ \Frac{d}{dt}\left(\Frac{\partial 
L}{\partial \dot{R}}\right)-\Frac{\partial L}{\partial R}=0$ 
gives the well known
relation $ 
R=6\left(\frac{\ddot{a}}{a}+\frac{\dot{a}^2}{a^2}\right)$,
while the second equation $ \Frac{d}{dt}\left(\Frac{\partial
L}{\partial \dot{a}}\right)- \Frac{\partial L}{\partial a}=0$ 
yields eq.~(\ref{H_dot}). For non-spatially flat FLRW 
universes
($k\ne 0$), the Hamiltonian constraint takes the form
\begin{equation}
H^2=\Frac{1}{3f'} 
\left[\Frac{Rf'-f}{2}-3H\dot{R}f''\right] -\Frac{k}{a^2}\; ,
\end{equation}
in which the scale factor $a$ appears explicitly, and one
can not use $H$ as variable instead of $a$. The surface $\Sigma$
studied for $k=0$ separates the region of the phase space accessible
to the orbits of the solutions for $k>0$ from the region
corresponding to $k<0$. The orbits can not cross the surface 
$\Sigma$ and, in
the case $k=0$, instead, they lie completely in $\Sigma$. A 
similar situation was recognized in an early study of inflation 
with a massive scalar field in the context of Einstein gravity 
\cite{Zeldovich}.

\subsection{The effective equation of state}

Using eqs.~(\ref{H_squa}) and (\ref{H_dot}) and identifying 
the effective gravitational coupling with $1/f'(R)$, one can 
write
the effective energy density and pressure for $f(R)$ models as 
follows:

\begin{equation}\label{density} 
\rho_{eff}= 
\frac{Rf'-f}{2}-3H\dot{R}f''  \;,
\end{equation}

\begin{equation}\label{pressure}
P_{eff}=  \dot{R²}f'''+  
2H\dot{R}f''+\ddot{R}f''+\frac{1}{2} \left( f-Rf' \right)  \;.
\end{equation}
The effective energy density $\rho_{eff}$ is non-negative, as 
can be seen from inspection of eq.~(\ref{H_squa}). The effective 
equation of state parameter $w_{eff}$ can be expressed as
\begin{equation}\label{eq_state}
w_{eff}  \equiv 
\Frac{P_{eff}}{\rho_{eff}}= 
\Frac{  \dot{R²}f'''+2H\dot{R}f''+ 
\ddot{R}f''+\Frac{1}{2}(f-Rf') }{ \Frac{Rf'-f}{2} 
-3H\dot{R}f'' } \;.
\end{equation}
Equation~(\ref{H_squa}) guarantees that the  denominator on the 
right hand side of eq.~(\ref{eq_state}) is strictly positive, 
hence, the 
sign of the effective equation of  state is determined by the 
numerator. Type Ia supernovae yield an effective equation of 
state $w\approx -1$ at present.  For the model to mimic  the de 
Sitter equation of state with $w_{eff}=-1$, it must be
\begin{equation}\label{ratio}
\Frac{f'''}{f''}=\Frac{ \dot{R}H -\ddot{R}}{( \dot{R})^2 }\;  .
\end{equation}
It is convenient to introduce the quantity  $\psi (R) \equiv 
f'(R)$ to write
\begin{equation}
w_{eff}= -1+2\, \frac{\left( \ddot{\psi}-H\dot{\psi} 
\right)}{R\psi-f-6H \dot{\psi} }= -1+\frac{\left( 
\ddot{\psi}-H\dot{\psi}\right) }{3\psi H^2} \;.
\end{equation}
Alternatively, the deviation from the de Sitter equation of 
state $w=-1$ can be parametrized by 
\begin{equation}  \label{questa}
\rho_{eff}+P_{eff}=\frac{\ddot{\psi}-H\dot{\psi}}{\psi}=\frac{\dot{\psi}}{\psi} 
\, \frac{d}{dt} \left[ \ln \left( \frac{\dot{\psi}}{a} \right) 
\right]\;.
\end{equation}
According to eq.~(\ref{questa}), an exact de  Sitter solution 
corresponds to $\dot{\psi}=f''(R) \dot{R}=0$, or to $\dot{\psi} 
=C a(t)=C a_0 \, \mbox{e}^{H_0 t}$, where $C\neq 0 $ is an 
integration constant. It is easy to see that the second 
solution for $\psi(t)$ is not 
acceptable because it leads to the absurd equation $f''(R) 
\dot{R}=C a_0\mbox{e}^{H_0 t}$ in which the left hand side is 
time-independent (for a de Sitter solution) while the right 
hand side is not.

\section{The equivalent scalar-tensor theory}\label{equiv_stt}
\setcounter{equation}{0}

A modified gravitational action of the form (\ref{action_1}) can 
be recast into the form of an equivalent scalar-tensor theory
(\cite{TeyssandierTourrenc,Wands}; see \cite{Whitt84} for a  
mapping into 
general relativity with a minimally coupled scalar). When 
$f''\neq 0$, the modified gravity action
(\ref{action_1}) can be rewritten as the scalar-tensor theory
\begin{equation}\label{action_2}
S=\frac{1}{2}\int d^4 x \, \sqrt{-g} \,[\psi(\phi)R-V(\phi)] 
+S^{(matter)} \;,
\end{equation}
where 
\begin{equation}\label{defs}
\psi(\phi)=f'(\phi) \;,\;\;\;\; \quad V(\phi)=\phi f'-f \;,
\end{equation}
and the Brans-Dicke parameter is $\omega=0$ 
\cite{TeyssandierTourrenc,Wands}. The Ricci curvature scalar $R$ is 
identified with the scalar field degree of freedom $\phi$ and 
the  mapping to scalar-tensor variables can be seen as a 
Legendre  transformation (see Ref.~\cite{MagnanoSokolowski} for 
a discussion, and Ref.~\cite{SO2} about the physical equivalence). In
fact, variation of the action (\ref{action_2}) with respect to $\phi$ yields
\begin{equation} \label{psiprime}
R \, \frac{d\psi}{d\phi}-\frac{dV}{d\phi}=0 \;,
\end{equation}
\noindent which in turn yields $\phi=R$ if $f''\neq 0$. There is no
kinetic term for the scalar $\phi$ in the action (\ref{action_2}), but
this quantity is dynamical because $R=\phi$ obeys the dynamical
equation (\ref{fieldeq}).

In the metric (\ref{metric}) the field equations become
\begin{equation}\label{H_squa_stt}
H^2=\frac{1}{6\psi} \left[ V(\phi)-6H\psi'\dot{\phi} \right] \; 
,
\end{equation}
\begin{equation}\label{H_dot_stt}
\dot{H}=\frac{1}{2\psi} \left( H\psi'\dot{\phi}- 
\psi'\ddot{\phi}-\psi''\dot{\phi^2} \right) \; ,
\end{equation}
and (\ref{psiprime}), or 
\begin{equation}\label{H_equiv} 
H^2=\frac{1}{6f'} \left( \phi f'-f-6Hf'''\dot{\phi} \right) \;, 
\end{equation} 
\begin{equation}\label{H_dot_equiv}
\dot{H}=\frac{1}{2f'} 
\left( Hf''\dot{\phi}-f''\ddot{\phi}-f'''\dot{\phi^2} 
\right)\; ,
\end{equation} 
\begin{equation}\label{f_twoprime} 
\left( R-\phi \right) f''=0 \;.  
\end{equation} 
By viewing the Hamiltonian constraint (\ref{H_squa_stt}) as an 
algebraic equation for $\dot{\phi}$, one can eliminate 
$\dot{\phi}$ from the three-dimensional phase space $ \left( H, 
\phi, \dot{\phi} \right) $ (which corresponds to the phase space 
$(H, R, \dot{R})$ of  Sec.~\ref{f_R}), obtaining 
\begin{equation}\label{phi_dot} 
\dot{\phi} \left( H, \phi \right) =\frac{1}{6Hf''}\left( \phi 
f'-f-6f'H^2 \right) \; .  
\end{equation}
This represents a one-sheeted energy surface, which corresponds 
to the special case $\omega=0$ of scalar-tensor cosmology
\cite{VFAnnPhys}, as remarked in the previous section. The 
equilibrium points $ \left( \dot{H}, \dot{\phi} \right) 
=\left( 0, 0 \right)$ are de Sitter spaces with
constant scalar field and satisfy
\begin{equation}\label{desitter_equiv}
\left( H_0, \phi_0 \right)=\left(\pm\sqrt\Frac{f_0}{6f'_0}, 
\Frac{2f_0}{f'_0}\right) \; ;
\end{equation}
they lie in the $\dot{\phi}=0$ plane. In general scalar-tensor 
gravity it is possible to have de Sitter solutions with 
non-constant scalar field (which, however, are not fixed 
points). In the scalar-tensor equivalent of $f(R)$ gravity this 
is not possible because of the constraint $\phi=R=12H_0^2 $ 
which forces $\phi$ to be constant for all de Sitter solutions, 
which have constant Hubble parameter $H_0$.

The gauge invariant stability condition for these de Sitter 
equilibrium points with 
respect to inhomogeneous perturbations, obtained in Ref. 
\cite{VFstability2}, is again
\begin{equation}
\frac{\left( f'_0 \right)^2-2f_{0}f''_{0}}{f'_{0}f''_0}\geqslant 
0 \;.
\end{equation}
The effective Lagrangian and Hamiltonian for the equivalent
scalar-tensor theory (\ref{action_2}) are now given by
\begin{equation}\label{lagrangian_stt}
L \left( a, \phi, \dot{a}, \dot{\phi} \right) = a^3 \left( 
6aH^2\psi+6H\dot{\psi}+V 
\right)
\end{equation}
and
\begin{equation}\label{hamiltonian_stt}
E \left( a, \phi, \dot{a}, \dot{\phi}  \right)=a^3 \left( 
6H^2\psi+6H\dot{\psi}-V \right) \; .
\end{equation}
These can be obtained from well known Lagrangians of
scalar-tensor theories (\cite{mybook} and references therein) 
and do not constitute an original result, contrary to the $L$ 
and $E$ given by eqs.~(\ref{lagrangian}) and 
(\ref{hamiltonian}).

\section{Discussion and conclusions}\label{conclusions}
\setcounter{equation}{0}

The phase space view of a spatially homogeneous and isotropic universe
is useful for building models of the current acceleration of our
cosmos and in trying to discover what fuels it. The high degree of
symmetry of the FLRW metric (\ref{metric}) considered turns the
Einstein equations into ordinary differential equations, for which the
Cauchy problem is trivial and is covered by well-known
theorems. Instead, the initial value problem of $f(R)$ gravity for
general metrics has not been discussed, except for specific choices of
the function $f(R)$ \cite{Noakes,TeyssandierTourrenc}, and will be
addressed elsewhere.

The dynamical system approach is a powerful tool which is appropriate
to study the dynamics for general initial conditions when exact
solutions can not be obtained (i.e., in most situations). As such,
dynamical systems theory is widely applied in cosmology (see, e.g.,
Refs.~\cite{WainwrightEllis, Coley}).  The theory of dynamical systems
has been applied repeatedly also to $f(R)$ cosmology, but usually for
special choices of the function $f(R)$ \cite{f(R)dynsyst}.  Our goal
is to study the phase space of $f(R)$ cosmology without committing to
a specific choice of the function $f(R)$.  Motivated by the recent
observations of type Ia supernovae \cite{SN} and by the current cosmic
microwave background experiments \cite{CMB}, we have restricted the
scope of our analysis to spatially flat FLRW cosmologies. Advantages
of our approach include its generality, the use of physical
observables as dynamical variables, and the ease of comparison with
the general picture of the phase space for the equivalent
scalar-tensor gravity. The obvious limitation is that the phase space
portrait is necessarily incomplete without choosing a specific form of
$f(R)$. However, in the absence of compelling physical indications on
the form of this function, it is hoped that our results on the
geometry of the phase space, the equilibrium points and their
stability, and the effective Lagrangians and Hamiltonians can be used
as a preliminary step and a guide to further understanding the
cosmological dynamics in any specific $f(R)$ theory. The extension to
more general gravity theories of the form $f \left( R, R_{ab}R^{ab},
R_{abcd}R^{abcd} \right)$ will be discussed elsewhere.

\ack{This work was supported by Conselho Nacional de Desenvolvimento
  Científico e Tecnológico (CNPq-Brazil), by the Natural
  Sciences and Engineering Research Council of Canada (NSERC), 
and by a Bishop's University Research Grant.}

\vskip1truecm


\end{document}